# Impact of exponential long range and Gaussian short range lateral connectivity on the distributed simulation of neural networks including up to 30 billion synapses

Elena Pastorelli[1], Pier Stanislao Paolucci[1*], Roberto Ammendola[2], Andrea Biagioni[1], Ottorino Frezza[1], Francesca Lo Cicero[1], Alessandro Lonardo[1], Michele Martinelli[1], Francesco Simula[1], Piero Vicini[1]

[1]INFN Roma "Sapienza", Italy
[2]INFN Roma "Tor Vergata", Italy

[*]Corresponding author: Pier Stanislao Paolucci, E-mail pier.paolucci@roma1.infn.it

**ABSTRACT**

Recent experimental neurosciences studies are pointing out the role of long-range intra-areal connectivity that can be modeled by a distance dependent exponential decay of the synaptic probability distribution. This short report provides a preliminary measure of the impact of exponentially decaying lateral connectivity compared to that of shorter-range Gaussian decays on the scaling behavior and memory occupation of a distributed spiking neural networks simulator (DPSNN). Two-dimensional grids of cortical columns composed by point-like spiking neurons have been connected by up 30 billion synapses using exponential and Gaussian connectivity models. Up to 1024 hardware cores, hosted on a 64 nodes server platform, executed the MPI processes composing the distributed simulator. The hardware platform was a cluster of IBM NX360 M5 16-core compute nodes, each one containing two Intel Xeon Haswell 8-core E5-2630 v3 processors, with a clock of 2.40GHz, interconnected through an InfiniBand network. This study is conducted in the framework of the CORTICONIC FET project, also in view of the next-to-start activities foreseen as part of the Human Brain Project (HBP), SubProject3 – Cognitive and Systems Neuroscience, WaveScalES work-package.

## 1. Introduction

Recent studies point out that remote connectivity plays a central role in many different cerebral areas, from cat primary visual cortex [1], to rat neocortex [2][3], just as examples. For instance, in rat neocortex, the impact of lateral connectivity on the pyramidal cells in layer 2/3 and layer 6A, results in ~75% of incoming remote synapses to neurons of these layers. The distance dependent slow decay of intra-areal non-local connection probability can be modeled by simple exponentials [2][3]. Decay constants in the range of several hundred microns has been proposed as matching the experimental results. This kind of intra-areal long-range lateral connectivity poses novel simulation requirements in comparison to previous studies that considered intra-areal synaptic connections dominated by local connectivity: for networks sufficiently large, local connections has been estimated as counting for at least 55% of total synapses, reaching also ratio of 75% [4]. In these studies, lateral connectivity has been often described with a Gaussian lateral connectivity model [5] reaching moderate distances, corresponding to a span of only a few cortical columns.
The increment in the range of remote connection is expected to have an impact on the performances of neural network simulators. On our DPSNN[1] simulator, the selection of the connectomic model is crucial, due to the fact that the synaptic messages exchanged between neurons correspond to communication tasks among MPI processes: the higher the number of lateral synaptic connections and the longer the distance is, the more intensive the communication task among processes becomes. The purpose of this study is to provide a first measure of this impact, comparing the scaling features of the DPSNN simulator implemented with connection matrices matching the two different connectivity models, for a preliminary assessment of the scalability of the code in case of longer range connectivity and growing number of synapses per neuron.

---

[1] The Distributed and Polychronous Spiking Neural Network simulator (DPSNN) with synaptic spike-timing dependent plasticity is a mixed time- and event-driven spiking neural network simulator designed to be natively distributed and parallel [6]. It has been initially developed in the framework of the EURETILE EU project (2010-2014) [7] as mini-application benchmark to characterize software and hardware architectures dedicated to neural simulations, and to drive the development of future generations of simulation systems (e.g. [8][9]). Later on, during the CORTICONIC EU project (2013-Mar 2016) [10], it has been improved including some of the main features of the scalar (not parallel, nor distributed) Perseo simulator [11], to become an efficient tool for large scale cortical simulations. Starting from April 2016, DPSNN will be used in the framework of WaveScalES, a part of the SP3 subproject of the Human Brain Project

## 2. Methods

The DPSNN is a spiking neural network simulator coded as a network of C++ processes, designed to be easily interfaced both to MPI and to custom Software/Hardware Communication Interfaces.
The simulator has been designed to be natively distributed and parallel. We address the reader to [6] for a more detailed description of the neural network simulator architecture, while [18] reports about the scaling behavior of DPSNN up to 1024 software processes (and hardware cores) for neural systems including up to 20G synapses.
The measures reported in this paper are relative to a configuration in which the adopted neuron model is the Leaky Integrate and Fire with spike-frequency adaptation (LIF with SFA) [19][20]. The neurons are organized in columns, each one composed of 80% excitatory and 20% inhibitory neurons. The columns are assembled in bi-dimensional grids, representing a cortical area slab, with a grid step $\alpha$ ~100 µm (columnar spacing). In order to study the scalability of the DPSNN simulator on large configurations, above all when connectivity between neurons changes, we considered two neural systems with different rules for the connectivity calculation. The local connectivity, i.e. the synapses generated by source neurons belonging to the same column of the target neuron, has been set to 80% in both cases. We adopted two different schemes for the lateral (remote) intra-areal connectivity, i.e. the synapses generated by neurons belonging to different columns placed at distance r. This produced two systems with different characteristics in terms of incoming synapses:

- Gaussian connectivity – shorter range and lower number of remote synapses: considering preeminent local connectivity with respect to lateral, the rule used to calculate decreasing remote connectivity has been set proportional to $A \cdot \exp(-r^2/2\sigma^2)$, with A=0.05 and $\sigma$=100 µm being respectively the peak amplitude and the lateral spread of the connection probability. In this way, the majority of connections are local (~79%), while the remaining (~21%) are remote and reach only columns placed within a short distance, spanning a few steps in the cortical column two-dimensional grid.
- Exponential connectivity – longer range and higher number of remote synapses: the connectivity rule for remote synapses calculation has been changed to be proportional to $A \cdot \exp(-r/\lambda)$, with A=0.03 and $\lambda$=290 µm being respectively the peak connection probability and the exponential decay constant. This turns out in an increased number of remote connections, as detailed in the following.

The number of neurons in each column has been fixed to 1240, while the number of synapses projected by each neuron depends on the kind of implemented connectivity: in case of lower connectivity, the average number of projected synapses per neuron is about 1240, while in case of higher connectivity this number rises up to ~2050 projected synapses per neuron. In both cases, the local synapses, i.e. the synapses project toward neurons of the same column, are about 990 (same local connectivity fixed at 80% in both systems), while the difference is in the remote connectivity: ~250 synapses for the first system, more than 1000 synapses for the second one.

---

In both systems, a cut-off has been set in the synapses generation, limiting the projection to the subset of columns with connection probability greater than 1/1000. This turns out in a centered stencil of connected columns of size 7x7 in the first case (Gaussian, lower connectivity) and 21x21 in the second case (exponential, higher connectivity).

In addition to the recurrent synapses, the system simulates also a number of external synapses: they represent afferent (thalamo-)cortical currents coming from outside the simulated network. The number of external synapses, firing at a rate of ~3Hz, has been fixed to 540 per neuron. The recurrent synapses plus the external synapses gives the number of total synapses afferent to the neuron, referred to as total equivalent synapses in the following.

For each of the two connectivity schemes, measurements has been taken on different problem sizes obtained varying the dimension of the grid of columns and, once fixed the problem size, distributing it over a span of MPI processes to evaluate the scaling behavior of the simulator. We selected three grid dimensions, which, e.g. for a columnar spacing of 100 μm, can be already considered representative of interesting biological cortical slab dimensions. Table 1 summarizes the main figures of the different problem sizes used in our simulations.

| Grid | Number of Columns | Number of Neurons | Number of Synapses | | | |
|---|---|---|---|---|---|---|
| | | | Gaussian shorter-range connectivity | | Exponential longer-range connectivity | |
| | | | Recurrent | Total | Recurrent | Total |
| **24x24** | 576 | 0.7M | 0.9G | 1.2G | 1.5G | 1.8G |
| **48x48** | 2304 | 2.9M | 3.5G | 5.0G | 5.9G | 7.4G |
| **96x96** | 9216 | 11.4M | 14.2G | 20.4G | 23.4G | 29.6G |

Table 1 – Problem sizes for the comparison of simulator performances applied to exponential (longer-range) and Gaussian (shorter-range) connectivity

The server platform used to run the simulations here described is GALILEO [21], a cluster of 516 IBM nodes provided by the consortium CINECA. Each 16-core computational node contains two Intel Xeon Haswell 8-core E5-2630 v3 processors, with a clock of 2.40GHz. All the nodes are interconnected through an InfiniBand network. Due to the specific configuration of the server platform, no hyper-threading is allowed. Therefore, in all the measures of this paper, the number of cores corresponds exactly to the number of MPI processes launched at each execution.

Using the Gaussian shorter-range connectivity, an extensive campaign of measures has been conducted, spanning on the three configurations described above. On the contrary, just a preliminary set of measures have been taken in the configuration with the longer range exponential connectivity: only few trials on 24x24 and on 48x48 configuration networks have been done.

The reason of the reduced number of measures taken for the higher connectivity case is in the much higher cost in core hours needed to run the higher connectivity simulations, paired with the limits imposed by the resources constrain assigned to this preliminary scaling investigation project. In the next period more extensive measures will be conducted, above all in the case of higher connectivity with exponential decay, in order to verify the preliminary results here reported.

Apart the probability of connection, all other simulation parameters (e.g. neuron dynamic, synaptic amplitudes and columnar grid sizes) have been kept identical among all the simulated configurations. The different connectivity produced different firing rates. The observed firing rate has been ~7.5 Hz for the lower connectivity scheme (the Gaussian decay), and in the range between 32 and 38 Hz for the higher interconnectivity scheme (the simple exponential decay).

## 3. Results

The execution times can been translated into a simulation cost per synaptic event dividing the elapsed time per simulated second of activity by the number of synapses and by the firing rates.

This way, a simple comparison among different simulated configurations is possible. In Figure 1, the strong scaling behavior of the DPSNN simulator in case of lower and higher lateral connectivity is reported for the set of two dimensional columnar grid dimensions. Circles represent measures for the Gaussian decay of connection probability (shorter range, lower number of synapses per neuron), while squares represents measures the exponential decay (longer range, higher number of synapses per neuron).

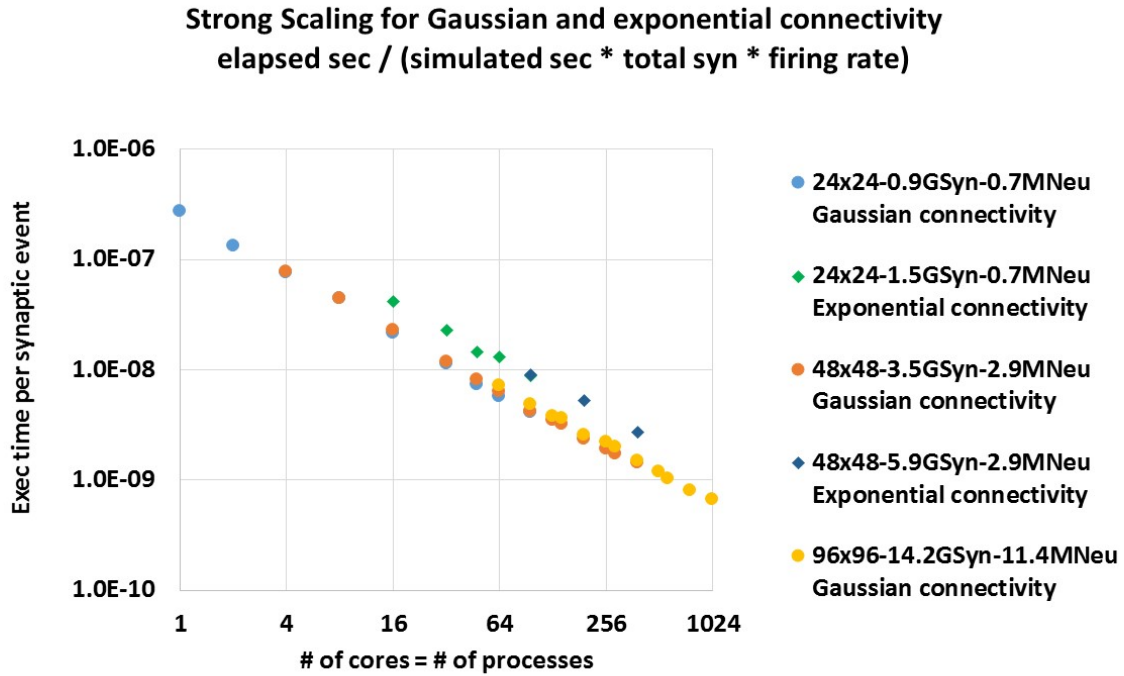

Figure 1. Strong scaling: configurations with Gaussian connectivity (shorter range, lower number of synapses - circles) and exponential connectivity (longer range, higher number - squares)

The impact of the simulation of the additional connectivity is indeed observable, as expected, and increases the simulation cost per synaptic event between 1.9 and 2.3 times (see also Figure 2). It is also worth nothing that: a) the actual elapsed simulation time increased up to 16.6 times for the exponential longer-range connectivity due to 1- the higher number of synapses projected by each neuron and 2- the higher firing rate of the exponentially connected network. We also remind that biologically accurate simulation would require the projection of a number of synapses in the range of ten thousand per neuron.

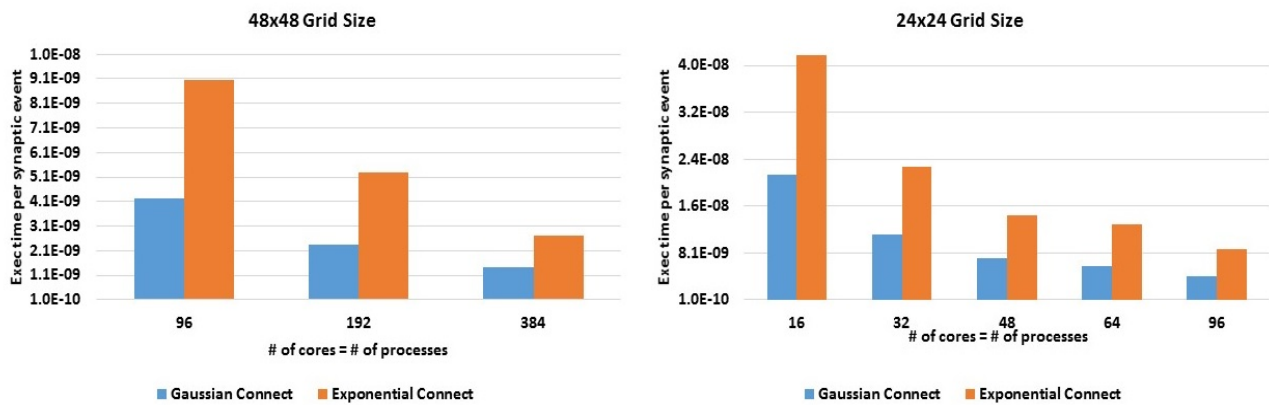

Figure 2. Time per simulated synaptic events increased between 1.9 and 2.3 times changing from the shorter range Gaussian scheme to the longer range exponential decrease of connection probability

The strong scaling shows interesting figures for both systems. For example, let us consider the case of 24x24 network with lower remote connectivity, implementing 0.9G recurrent synapses and 1.2G total equivalent synapses, with a total of 0.7M neurons: with 1 single process, 2.8E-7 seconds are required to simulate a synaptic event, that reduce to 4.2E-09 seconds when using 96 processes. This means that increasing the number of processes by 96, the execution time speeds-up of a factor 66.7, , i.e. about the 70% of the ideal scaling case. Similar numbers arise analyzing the other two configurations, 48x48 (57% of ideal case) and 96x96 (68%), always remaining in the case of the shorter range remote connectivity.

Doing the same analysis on the two configurations with longer range remote connectivity, we figure out better results, with about 79% of the ideal case for the 24x24 (1.5G recurrent synapses) and 83% for the 48x48 (5.9G recurrent synapses).

About memory occupation, the number of byte per synapse required by the simulation is comparable in the two systems with lower and higher connectivity, also varying the size of the tested configuration, as reported in Figure 3.

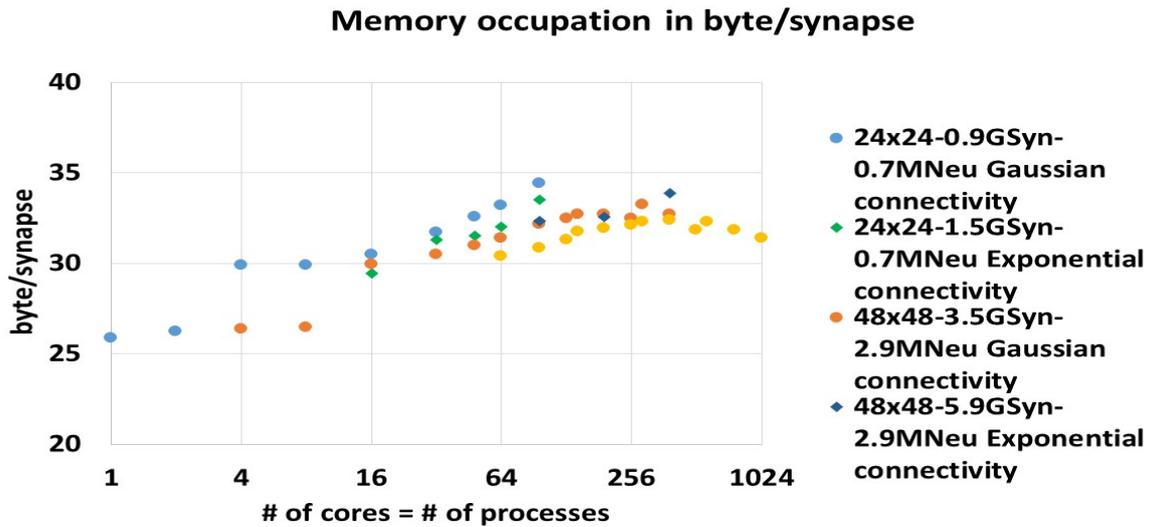

Figure 3. Memory occupation in byte per synapse for different configurations in the two connectivity systems

## 4. Discussion

As discussed in the Introduction, recent experimental results suggest the need of supporting long range lateral connectivity in neural simulation of cortical areas, e.g. modeled by simple exponential decay of the connection probability, with layer to layer specific decay constants, in the order of several hundreds micron.

The distributed spiking neural net simulator DPSNN has been applied to two-dimensional grids of neural columns spaced at 100 μm connected using two schemes. The longer-range connectivity model corresponds to an exponential connectivity decay ($\lambda$=290 μm), and to the projection of approximately ~2050 synapses per neuron. The scaling measures are compared to those obtained with a shorter range Gaussian decay of the connectivity, with a decay constant of the order of the columnar spacing and a lower number of synapses per neuron (~1240).

The impact of exponentially decaying connectivity is indeed observable, as expected, and increases the simulation cost per synaptic event between 1.9 and 2.3 times compared to a shorter range Gaussian connectivity law.

Notwithstanding this increase, the strong scaling behavior is still satisfactory.

However, we note that a more realistic biological simulation could require a further extension of the connection stencil dimension, with the goal of projecting about ten thousand synapses per neuron. A further significant increase of the simulation cost per synaptic event is expected.

For what concern memory, the byte per synapse required by the simulation of networks with the same number of neurons, but different number of synapses per neuron, is comparable.

## 5. Acknowledgements

The DPSNN simulator development has been partially funded by the EURETILE European project (FET grant no. 247846) and by the CORTICONIC (FET grant no. 600806) European projects.
We acknowledge the key advices of Paolo Del Giudice and Maurizio Mattia, in the framework of the CORTICONIC project.
Further, we thank Leonardo Cosmai for his support in running simulations on the Galileo supercomputer at CINECA.